\documentclass[aps,prl,twocolumn,amsfonts,amsmath,floatfix,showpacs,superscriptaddress,footnoteintext]{revtex4}

\usepackage[final]{graphicx}
\usepackage{color}

\newcommand{\be}{\begin{equation}} \newcommand{\ee}{\end{equation}}

\begin{document}

\title{Wave-scattering  from a gently curved  surface}

\author{Giuseppe Bimonte}

\affiliation{Dipartimento di Fisica E. Pancini, Universit\`{a} di
Napoli Federico II, Complesso Universitario
di Monte S. Angelo,  Via Cintia, I-80126 Napoli, Italy}
\affiliation{INFN Sezione di Napoli, I-80126 Napoli, Italy}

\begin{abstract}
{We study wave scattering from a gently curved surface. We show that the recursive relations, implied by shift invariance,  among the coefficients of the perturbative series for the scattering amplitude allow to perform an infinite resummation of the perturbative series  to all orders in the amplitude of the corrugation.  The resummed series provides a derivative expansion of the scattering amplitude in powers of derivatives of the height profile, which is expected to become exact   in the limit of quasi-specular scattering. We discuss the relation of our results with the so-called small-slope approximation introduced some time ago by Voronovich. }
\end{abstract}

\maketitle

\section{Introduction} 

The problem of wave scattering  at rough interfaces has always attracted much attention, due to its importance in diverse areas of physics ranging from optics, to acoustics, communications, geophysics etc. Its difficulty is univerally acknowledged, and even with presently available computers  it represents a formidable challange in realistic  situations. Despite impressive progress made in numerical methods, approximate analytical approaches are still much valuable as  the only ones  capable of providing  an  insight into general physical features of wave scattering.  It is not surprising then that a number of diverse approximation schemes, with different ranges of validity, have been developed over the years. For a review, we address the reader to the monograph \cite{voron2} or to the recent review \cite{guerin}. Hystorically, the first and perhaps most widely known approximate theory of scattering by rough surfaces is the {\it small perturbation method} (SPM) originally developed by Rayleigh to study scattering of sound waves by sinusoidally corrugates surfaces of small amplitude, and later generalized by several other authors to electromagnetic scattering. Another classic approximate scheme is the {\it Kirchhoff approximation} (KA), also known as the {\it tangent plane} approximation (TPA), which represents a valid approximation for locally smooth surfaces with large radii of curvature. A  scheme  aiming at reconciling SPM and KA was proposed twenty years ago by Voronovich \cite{voron2,voron}, i.e. the {\it small-slope approximation} (SSA). It consists of an ingenious {\it ansatz} for the scattering amplitude (SA),  that manages to capture the leading curvature correction to the KA. The ansatz involves unknown coefficients that are determined {\it a posteriori} by matching the SSA with the SPM in their common region of validity.  Numerical investigations revealed that within its domain of validity the SSA is indeed remarkably accurate \cite{guerin}. Several authors have attempted to provide rigorous mathematical derivations of the SSA, utilizing the extinction theorem \cite{tatar1,tatar2} or the Meecham-Lysanov method \cite{McD}. 

In recent years it has been shown \cite{fosco0,bimonte1,bimonte2} that curvature corrections to the Casimir interaction between two  gently curved surfaces can be estimated using a derivative expansion (DE) of the Casimir-energy functional, in powers of derivatives of the height profiles of the surfaces. The DE has been later used to study curvature effects in the Casimir-Polder interaction of a particle  with a gently curved surface \cite{bimonte3,bimonte4}. The same method has been used very recently to estimate the shifts of the rotational levels of a diatomic molecule due to its van der Waals interaction with a curved dielectric surface \cite{bimonte5}.  An important insight into the nature of the  DE was achieved in \cite{fosco} (see also \cite{bimonte3}) where it was shown that the derivative expansion  amounts to an infinite  resummation of the perturbative series for the Casimir energy to all orders in the amplitude of the corrugation, for {\it small} in-plane momenta of the electromagnetic field. These methods can indeed be adapted to the problem of wave scattering by a rough surface.     
In this letter we demonstrate that the infinite recursive relations among the coefficients of the perturbative series for the SA, engendered by its exact {invariance} under vertical and hortizontal shifts of the surface,  allow to perform an {\it infinite} resummation of the perturbative series, 
to all orders in the amplitude of the corrugation. The resummation procedure results into a  DE of the  SA in powers of derivatives of the height profile, which is expected to become exact   in the limit of quasi-specular scattering.  We show that our  DE of the SA is indeed equivalent to Voronovich's SSA ansatz, thus providing a formal justification for it.

\section{Perturbative expansion of the SA}

We consider a surface $\Sigma$ separating two media of different (optical or acoustical) properties. A cartesian coordinate system $(x,y,z)$ is chosen such that the $z$-axis is directed upwards in the direction going from medium 2 towards medium 1, while $(x,y)$ span a reference plane orthogonal to the $z$-axis. It is assumed that the surface $\Sigma$  can be represented by a (single-valued) {\it smooth} height profile of equation $z=h({\bf x})=h(x,y)$. A downward propagating (electromagnetic or acoustic)  wave ${ E}^{(\rm in)}_{\alpha_0}({\bf x},z)= {E}_{\alpha_0}^{(\rm 0)}/\sqrt{q_0}\;\exp[{\rm i}({\bf k}_0 \cdot {\bf x}-q_0 z)]$ with wave vector ${\bf K}_0=({\bf k}_0,-q_0)$,  wavenumber $K_0=2 \pi/\lambda_0$  \footnote{We follow the normalization of waves adopted by Voronovich \cite{voron}.} is incident on the surface. The index $ \alpha_0$ takes values in a discrete set consisting of one or two elements, depending on whether an acoustic or electromagnetic wave is considered. The scattered field ${\bf E}^{(\rm sc)}({\bf r},z)$ at points  above and far from $\Sigma$ can be expressed  through the  SA ${S}_{\alpha \alpha_0}({\bf k},{\bf k}_0)$ as a superpositions of upwards propagating waves  with wave vector ${\bf K}=({\bf k},q)$  as:  
\be
{ E}^{(\rm sc)}_{\alpha}({\bf x},z)=\int \frac{d^2 {\bf k}}{4 \pi^2}\frac{1}{\sqrt{q}}\,e^{{\rm i}({\bf k} \cdot {\bf x}+q z)}\,{S}_{\alpha \alpha_0}({\bf k},{\bf k}_0)\,{E}_{\alpha_0}^{(\rm 0)}\;.
\ee
The SA satisfies two general properties. The first one is {\it reciprocity}:
\be
{S}_{\alpha \alpha_0}({\bf k},{\bf k}_0)={S}_{\alpha_0 \alpha}(-{\bf k}_0,-{\bf k})\;,
\ee
which follows from microscopic reversibility \cite{landau}. The second general property satisfied by the SA is {\it shift invariance}, which  amounts to the following transformation property of the SA under a horizontal and vertical displacement of the surface $\Sigma$: 
\be
S_{\alpha \alpha_0}({\bf k},{\bf k}_0 )\vert_{h({\bf x}-{\bf a})-b}  = e^{{\rm i} [({\bf k}_0-{\bf k})\cdot {\bf a} -(q_0+q) {b}]}S_{\alpha \alpha_0}({\bf k},{\bf k}_0)\vert_{h({\bf x})}\;.\label{tra1}
\ee  
According to the SPM, we postulate that for sufficiently small height profiles $h$ the SA can be expanded as a power series in the height profile: 
$$
S_{\alpha \alpha_0}({\bf k},{\bf k}_0)=\sum_{n \ge 0}\frac{1}{n!}\int d^2 {\bf x}_1 \cdots \int d^2 {\bf x}_n 
\;  
$$
\be
\times \,
G_{\alpha \alpha_0}^{(n)}(  {\bf x}_1,\dots,  {\bf x}_n;{\bf k},{\bf k}_0)\;
\,h(  {\bf x}_1) \dots h(  {\bf x}_n)\;,\label{pertx}
\ee
where  the  kernels $G^{(n)}_{\mu \nu}(  {\bf x}_1,\dots,  {\bf x}_n;{\bf k},{\bf k}_0)$ are  symmetric functions of $({\bf x}_1\,\dots {\bf x}_n)$.
In momentum space, the perturbative expansion of the SA  reads:
$$
S_{\alpha \alpha_0}({\bf k},{\bf k}_0)=\sum_{n \ge 0}\frac{1}{n!}\int \frac{d^2 {\bf k}_1}{(2 \pi)^2} \cdots \int  \frac{d^2 {\bf k}_n}{(2 \pi)^2} 
\;  
$$
\be
\times\; {\bar G}_{\alpha \alpha_0}^{(n)}(  {\bf k}_1,\dots,  {\bf k}_n;{\bf k},{\bf k}_0)\;{\tilde h}(  {\bf k}_1) \dots {\tilde h}(  {\bf k}_n)\;.\label{foupert}
\ee
Shift invariance under a {\it horizontal} translation of the profile $h({\bf x})$ implies  that the  kernels ${\tilde G}_{\alpha \alpha_0}^{(n)}(  {\bf k}_1,\dots,  {\bf k}_n;{\bf k},{\bf k}_0)$ must be of the form:
$$
{\bar G}_{\alpha \alpha_0}^{(n)}(  {\bf k}_1,\dots,  {\bf k}_n;{\bf k},{\bf k}_0)=(2 \pi)^2 \delta^{(2)}({\bf k}_1+\dots {\bf k}_n+{\bf k}_0-{\bf k})
$$
\be
\times
 \, {\tilde G}_{\alpha \alpha_0}^{(n)}(  {\bf k}_1,\dots,  {\bf k}_n;{\bf k},{\bf k}_0)\;,\label{gdelta}
\ee
where ${\tilde G}_{\alpha \alpha_0}^{(n)}(  {\bf k}_1,\cdots,  {\bf k}_n;{\bf k},{\bf k}_0)$ are symmetric functions of the in-plane momenta ${\bf k}_1,\dots,{\bf k}_2$,  which are defined only on the hyperplane ${\cal P}^{(n)} \equiv \{{\bf k}_1+\dots {\bf k}_n+{\bf k}_0-{\bf k}=0 \}$. Of course
\be
{\bar G}_{\alpha \alpha_0}^{(0)}({\bf k},{\bf k}_0)=(2 \pi)^2 \delta^{(2)}({\bf k}_0-{\bf k}){R}_{\alpha \alpha_0}({\bf k}_0)\;,
\ee 
where ${R}_{\alpha \alpha_0}({\bf k}_0)$ are the familiar  reflection coefficients for a planar surface.  By inserting Eq. (\ref{gdelta}) into Eq. (\ref{foupert}), the perturbative series can be rewritten as:
$$
S_{\alpha \alpha_0}({\bf k},{\bf k}_0)=\sum_{n \ge 0}\frac{1}{n!}\int \frac{d^2 {\bf k}_1}{(2 \pi)^2} \cdots \int  \frac{d^2 {\bf k}_n}{(2 \pi)^2}\, \,{\tilde h}(  {\bf k}_1) \dots {\tilde h}(  {\bf k}_n)
$$
\be
\times (2 \pi)^2 \delta^{(2)}({\bf k}_1+\dots {\bf k}_n+{\bf k}_0-{\bf k}){\tilde G}_{\alpha \alpha_0}^{(n)}(  {\bf k}_1,\dots,  {\bf k}_n;{\bf k},{\bf k}_0)
\; \;.\label{pert}
\ee
Next we  show that the perturbative kernels have to satisfy an infinite set of relations, as a result of the shift invariance under a {\it vertical} shift of the profile. To derive these relations we note the identity that follows from Eq. (\ref{tra1}):
\be
e^{{\rm i} (q_0+q) {b} }\;S_{\alpha \alpha_0}({\bf k},{\bf k}_0)\vert_{h({\bf x})-b}=S_{\alpha \alpha_0}({\bf k},{\bf k}_0)\vert_{h({\bf x})}\;.
\ee 
Upon taking  $p$ derivatives of both sides of the above relation with respect to the shift $b$, we obtain the identities:
$$
 \left.\frac{d^p}{d\,b^p}\left(e^{{\rm i} (q_0+q) {b} }S_{\alpha \alpha_0}({\bf k},{\bf k}_0)\vert_{h({\bf x})-b} \right)\right \vert_{b=0}
$$
\be
\equiv A^{(p)}_{\alpha \alpha_0}({\bf k},{\bf k}_0)\vert_{h({\bf x})}=0  \label{relas}
\ee
that have to be satisfied for {\it any} profile $h$. 
By making use of the perturbative expansion of the SA Eq. (\ref{pertx})   into the l.h.s. of the above identities,   one obtains the following expansion for the  kernels $A^{(p)}_{\alpha \alpha_0}({\bf k},{\bf k}_0)\vert_{h({\bf x})}$ :
$$
A^{(p)}_{\alpha \alpha_0}({\bf k},{\bf k}_0)\vert_{h({\bf x})}= \sum_{n \ge 0}\frac{1}{n!}\int d^2 {\bf x}_1 \cdots \int d^2 {\bf x}_n 
$$
\be
\times
\; A_{\alpha \alpha_0}^{(p,n)}(  {\bf x}_1,\dots,  {\bf x}_n;{\bf k},{\bf k}_0)\; h(  {\bf x}_1) \dots h(  {\bf x}_n)=0\;,
\ee
with:
$$
A_{\alpha \alpha_0}^{(p,n)}(  {\bf x}_1,\dots,  {\bf x}_n;{\bf k},{\bf k}_0) =\sum_{k=0}^p (-1)^k \frac{p!}{k! (p-k)!} \,[{\rm i} \,(q_0+q)]^{p-k} 
$$
\be
\times
\;\int d^2 {\bf x}_{n+1} \dots \int d^2 {\bf x}_{n+k} 
\; G_{\alpha \alpha_0}^{(n+k)}(  {\bf x}_1,\dots,  {\bf x}_{n+k};{\bf k},{\bf k}_0)\;,
\ee
Since the  identities in Eq. (\ref{relas}) must be satisfied for arbitrary profiles $h$, it follows that all the kernels $ A_{\alpha \alpha_0}^{(p,n)}(  {\bf x}_1,\dots,  {\bf x}_n;{\bf k},{\bf k}_0)$ must vanish identically:
\be
A_{\alpha \alpha_0}^{(p,n)}(  {\bf x}_1,\dots,  {\bf x}_n;{\bf k},{\bf k}_0)=0\;,\;\;\;\forall\; n \ge 0\;,p>0\;.
\ee
When expressed in momentum space, these conditions read:
$$
 (2 \pi)^2 \delta^{(2)}({\bf k}_1+\dots {\bf k}_n+{\bf k}_0-{\bf k})\;\sum_{k=0}^p (-1)^k \frac{p!}{k! (p-k)!}\;  
$$
\be
\times [{\rm i} (q_0+q)]^{p-k} \, {\tilde G}_{\alpha \alpha_0}^{(n+k)}(  {\bf k}_1,\dots,  {\bf k}_{n},{\bf 0},\dots,{\bf 0})=0\;.
\ee
For any fixed $n$,  the above relations can be solved iteratively leading to:
$$
 {\tilde G}_{\alpha \alpha_0}^{(n+m)}(  {\bf k}_1,\dots,  {\bf k}_{n},{\bf 0},\dots,{\bf 0};{\bf k},{\bf k}_0)
$$
\be
=[{\rm i} (q_0+q)]^{m}\;{\tilde G}_{\alpha \alpha_0}^{(n)}(  {\bf k}_1,\dots,  {\bf k}_{n};{\bf k},{\bf k}_0)\;.\label{rel}
\ee
We recall that these identities hold  on ${\cal P}^{(n)} $. Equations (\ref{rel}) constitute  a very important result, and in next Section we  show that with their help it is  possible to perform an {\it infinite} re-summmation of the perturbative series, order by order  in a small  ${\bf k}$ expansion. 

\section{Resummation of the perturbative series}

Now we consider a surface $\Sigma$ whose  local radius of curvature is everywhere large compared to the wavelength of the incoming wave.  
For such a gently curved profile, the Fourier transform ${\tilde h}({\bf k})$ of the height profile is significantly different from zero only for small in-plane wave vectors ${\bf k}$.  For quasi specular-scattering, i.e. for ${\bf k} \simeq {\bf k}_0$, it is therefore legitimate to  Taylor-expand the kernels  ${\tilde G}_{\alpha \alpha_0}^{(n)}(  {\bf k}_1,\dots,  {\bf k}_{n};{\bf k},{\bf k}_0)$   around ${\bf k}_1=\cdots={\bf k}_n={\bf 0}$.  Existence of this Taylor expansion  is not guaranteed a priori,  but has to be checked case by case.  With respect to this question, it is important to observe that by virtue of the iterative relations Eqs. (\ref{rel})  existence of the Taylor expansion of order $n$ for the  kernel ${\tilde G}_{\alpha \alpha_0}^{(n)}(  {\bf k}_1,\dots,  {\bf k}_{n};{\bf k},{\bf k}_0)$ automatically implies existence of the Taylor expansion  to the same  order $n$  for all higher order   kernels ${\tilde G}_{\alpha \alpha_0}^{(n+k)}(  {\bf k}_1,\dots,  {\bf k}_{n+k};{\bf k},{\bf k}_0)$, $k=1,2,\dots$.   Let us assume that the  perturbative kernels can be Taylor-expanded to second order in the wave vectors \footnote{By an explicit perturbative computation to second order in the height profile, it is possible to verify that the perturbative kernels   ${\tilde G}_{\alpha \alpha_0}^{(1)}(  {\bf k}_1;{\bf k},{\bf k}_0)$ and ${\tilde G}_{\alpha \alpha_0}^{(2)}(  {\bf k}_1,  {\bf k}_2;{\bf k},{\bf k}_0) $ do admit a second-order Taylor-expansion as per Eq. (\ref{exp2}), for both  a scalar wave satisfying Dirichlet or Neumann boundary conditions on the surface $\Sigma$, as well as for the scattering of an electromagnetic wave by a dielectric surface. Explicit formulae for the relevant perturbative kernels in the electromagnetic case are provided in \cite{voron}.}:
$$
{\tilde G}^{(n)}(  {\bf k}_1,\dots,  {\bf k}_{n};{\bf k},{\bf k}_0)
$$
$$=A^{(n)}({\bf k},{\bf k}_0)+B^{(n)}_{\mu}({\bf k},{\bf k}_0) ({\bf k}_1+\dots+  {\bf k}_{n})^{\mu}
$$
\be
+C^{(n)}_{\mu \nu}({\bf k},{\bf k}_0) \sum_{i=1}^n  {\bf k}_{i}^{\mu}  {\bf k}_{i}^{\nu}-D^{(n)}_{\mu \nu}({\bf k},{\bf k}_0) \sum_{i<j}  {\bf k}_{i}^{\mu}  {\bf k}_{j}^{\nu}+ {\rm o}(k^2) \;,\label{exp2}
\ee
where greek indices take values (1,2),   and for brevity we suppressed all polarization indices. Since the n-point Green function is defined only on the hyperplanes ${\cal P}^{(n)}$, we are free to add to ${\tilde G}^{(n)}(  {\bf k}_1,\cdots,  {\bf k}_{n};{\bf k},{\bf k}_0)$  any function $f^{(n)}( {\bf k}_1,\cdots,  {\bf k}_{n};{\bf k},{\bf k}_0)$ vanishing on ${\cal P}^{(n)}$ of the form 
$$
f^{(n)}( {\bf k}_1,\dots,  {\bf k}_{n};{\bf k},{\bf k}_0)=({\bf k}_1+\dots {\bf k}_n+{\bf k}_0-{\bf k})^{\mu}\;
$$
\be
\times\,g^{(n)}_{\mu}( {\bf k}_1,\dots,  {\bf k}_{n};{\bf k},{\bf k}_0)
\ee
where $g^{(n)}_{\mu}( {\bf k}_1,\cdots,  {\bf k}_{n};{\bf k},{\bf k}_0)$ are arbitrary  smooth symmetric functions of   $ {\bf k}_1,\cdots,  {\bf k}_{n}$. Let us take
\be
g^{(n)}_{\mu}={\hat A}^{(n)}_{\mu}({\bf k},{\bf k}_0)+{\hat B}^{(n)}_{\mu \nu}({\bf k},{\bf k}_0) ({\bf k}_1+\dots+  {\bf k}_{n})^{\nu}.
\ee
It is easy to verify that the coefficients ${\hat A}^{(n)}_{\mu}, \;{\hat B}^{(n)}_{\mu \nu}$ can always  be chosen such that  additon of $f^{(n)}( {\bf k}_1,\cdots,  {\bf k}_{n};{\bf k},{\bf k}_0)$ to  ${\tilde G}^{(n)}(  {\bf k}_1,\cdots,  {\bf k}_{n};{\bf k},{\bf k}_0)$  removes from Eq. (\ref{exp2}) the terms proportional to  ${ B}^{(n)}_{\mu}$ and  ${ C}^{(n)}_{\mu \nu}$. Without loss of generality, the  second order Taylor expansion of the kernels  ${\tilde G}^{(n)}(  {\bf k}_1,\cdots,  {\bf k}_{n};{\bf k},{\bf k}_0)$ around ${\bf k}_1=\cdots={\bf k}_n={\bf 0}$  can thus be assumed to be of the form:
$$
{\tilde G}^{(n)}(  {\bf k}_1,\cdots,  {\bf k}_{n};{\bf k},{\bf k}_0)
$$
\be
=A^{(n)}({\bf k},{\bf k}_0) -D^{(n)}_{\mu \nu}({\bf k},{\bf k}_0) \sum_{i<j}  {\bf k}_{i}^{\mu}  {\bf k}_{j}^{\nu}+ {\rm o}(k^2)   \;.\label{exp2bis}
\ee
After    the  above small-${\bf k}$ expansion is substituted into the perturbative series Eq. (\ref{pert}), one finds:
$$
S({\bf k},{\bf k}_0)=\sum_{n \ge 0}\frac{1}{n!}\int \frac{d^2 {\bf k}_1}{(2 \pi)^2} \cdots \int  \frac{d^2 {\bf k}_n}{(2 \pi)^2}  \; {\tilde h}(  {\bf k}_1) \dots {\tilde h}(  {\bf k}_n)\,\times
$$
$$
(2 \pi)^2 \delta^{(2)}({\bf k}_1+\dots {\bf k}_n+{\bf k}_0-{\bf k})
[A^{(n)}({\bf k},{\bf k}_0) -D^{(n)}_{\mu \nu}({\bf k},{\bf k}_0) \sum_{i<j}  {\bf k}_{i}^{\mu}  {\bf k}_{j}^{\nu}]
$$
$$
= (2 \pi)^2 \delta^{(2)}({\bf k}_0-{\bf k}){R}({\bf k}_0)+ A^{(1)}({\bf k},{\bf k}_0){\tilde h}(  {\bf k}- {\bf k}_0)
$$
$$
+\sum_{n \ge 2} \int d^2{\bf x} \;e^{{\rm i}({\bf k}_0-{\bf k}) \cdot {\bf x}}\left[ \frac{1}{n!}A^{(n)}({\bf k},{\bf k}_0)\,h^n({\bf x}) \right.
$$
\be
\left.+\frac{h^{n-2}({\bf x})}{2(n-2)!}D^{(n)}_{\mu \nu}({\bf k},{\bf k}_0) \partial_{\mu}h({\bf x})
 \partial_{\nu}h({\bf x}) \right]\;.\label{pertbis}
\ee
However, the identities Eq. (\ref{rel}) satisfied by the perturbative kernels for $n=0,1,2$  imply at once:
\be
 A^{(m)}({\bf k}_0,{\bf k}_0)=(2\,{\rm i}\, q_0)^{m}\ R({\bf k}_0)\;,\label{rec1}
\ee 
\be
 A^{(m+1)}({\bf k},{\bf k}_0)= [{\rm i} (q_0+q)]^{m}\ A^{(1)}({\bf k},{\bf k}_0)\;,\label{rec2}
\ee
\be
 A^{(m+2)}({\bf k},{\bf k}_0)= [{\rm i} (q_0+q)]^{m}\ A^{(2)}({\bf k},{\bf k}_0)\;,
\ee
\be
D^{(m+2)}_{\mu \nu}({\bf k},{\bf k}_0) =[{\rm i} (q_0+q)]^{m}\  D^{(2)}_{\mu \nu}({\bf k},{\bf k}_0) \;.\label{rec4}
\ee 
By making use into  Eq.(\ref{pertbis}) of the above identitites,  it is easy  to perform the infinite sums  in the r.h.s. of  Eq.(\ref{pertbis}): 
$$
S({\bf k},{\bf k}_0)= (2 \pi)^2 \delta^{(2)}({\bf k}_0-{\bf k})\left[{R}({\bf k}_0)-\frac{A^{(1)}({\bf k}_0,{\bf k}_0)}{2 {\rm i} q_0}\right]
$$
$$+ \int d^2{\bf x} \;e^{{\rm i}({\bf k}_0-{\bf k}) \cdot {\bf x}}\left\{\left[ \frac{A^{(1)}({\bf k},{\bf k}_0)}{{\rm i}(q_0+q)} \right.\right.
$$
$$
 \left.+\frac{1}{2} D^{(2)}_{\mu \nu}({\bf k},{\bf k}_0) \partial_{\mu}h({\bf x})
 \partial_{\nu}h({\bf x})\right] \left.\sum_{n \ge 0}\frac{1}{n!}\,[{\rm i}(q+q_0)h({\bf x})]^n \right\}
$$
$$
= \int d^2{\bf x} \;e^{{\rm i}[({\bf k}_0-{\bf k}) \cdot {\bf x}+(q+q_0)h({\bf x})]}\left[\frac{A^{(1)}({\bf k},{\bf k}_0)}{{\rm i}(q_0+q)} \right.
$$
\be
+\left.\frac{1}{2} D^{(2)}_{\mu \nu}({\bf k},{\bf k}_0) \;\partial_{\mu}h({\bf x})
 \partial_{\nu}h({\bf x})\right] \;.\label{ssa2}
\ee
The above  formula for the SA represents our main result. 
As we see, the contribution proportional to $A^{(1)}({\bf k},{\bf k}_0)$ reproduces the KA, 
while the second term of order $(\nabla h)^2$ 
provides the leading curvature correction to the KA.  Bearing in mind Eq. (\ref{exp2bis}), we see that the coefficients $A^{(1)}({\bf k},{\bf k}_0)$ and $ D^{(2)}_{\mu \nu}({\bf k},{\bf k}_0)$ occurring in this formula can be extracted from the second order Taylor expansion  of the perturbative kernels ${\tilde G}_{\alpha \alpha_0}^{(1)}(  {\bf k}_1;{\bf k},{\bf k}_0)$ and ${\tilde G}_{\alpha \alpha_0}^{(2)}(  {\bf k}_1,  {\bf k}_2;{\bf k},{\bf k}_0) $, respectively.  We now prove that our  Eq. (\ref{ssa2}) is indeed equivalent (to order $(\nabla h)^2$) to the SSA ansatz made by Voronovich \cite{voron} twenty years ago. 

\section{Comparison with Voronovich's small-slope approximation.}

In Ref. \cite{voron} Voronovich postulated the following {\it ansatz} for the SSA  valid to order $(\nabla h)^2$:
$$
S({\bf k},{\bf k}_0)=\frac{2\sqrt{q q_0}}{q+q_0}\int d^2{\bf x} \;e^{{\rm i}[({\bf k}_0-{\bf k}) \cdot {\bf x}+(q+q_0)h({\bf x})]}
$$
\be
\times \left[B^{11} ({\bf k},{\bf k}_0)-\frac{{\rm i}}{4}
\int d^2 {\bf k}_1 \;M^{11}({\bf k},{\bf k}_0;{\bf k}_1) {\tilde h}({\bf k}_1)\; e^{{\rm i} {\bf k}_1 \cdot {\bf x}}\right]\;,\label{ssavor}
\ee
where 
$$
M^{11}({\bf k},{\bf k}_0;{\bf k}_1)=B^{11}_2({\bf k},{\bf k}_0;{\bf k}-{\bf k}_1)
$$
\be
+B^{11}_2({\bf k},{\bf k}_0;{\bf k}_0+{\bf k}_1)+2 (q+q_0)\:B^{11}({\bf k},{\bf k}_0)\;,
\ee
and for brevity we suppressed all polarization indices. The perturbative kernels occurring in the above Equations  are related to ours by the following relations:
\be
{\tilde G}^{(1)} ({\bf k}-{\bf k}_0;{\bf k},{\bf k}_0)=A^{(1)}({\bf k},{\bf k}_0)=2 {\rm i}\,\sqrt{q q_0} B^{11} ({\bf k},{\bf k}_0)\;,\label{comp1}
\ee
and
$$
{\tilde G} ^{(n+1)}(  {\bf k}-{\bf k}_1, {\bf k}_2, \dots,  {\bf k}_n,  {\bf k}_n-{\bf k}_0;{\bf k},{\bf k}_0)
$$
\be
=(n+1)!\;\sqrt{q q_0}\; (B_{n+1})^{11}({\bf k},{\bf k}_0;
{\bf k}_1, \dots,  {\bf k}_n)\;,\;\;\;\;n=1,2,\dots  \label{comp2}
\ee

In view of of Eq. (\ref{comp1}), it is clear that the first term between the square brackets in Eq. (\ref{ssavor}) reproduces the first  term between the square brackets of Eq. (\ref{ssa2}).  By an explicit computation, it is possible to verify that the second terms between the square brackets of Eqs. (\ref{ssavor}) and (\ref{ssa2}) coincide as well. In view of Eq. (\ref{comp2}), and using the second order Taylor expansion of  ${\tilde G}^{(2)}( {\bf k}_1, {\bf k}_2;{\bf k},{\bf k}_0)$ given in Eq. (\ref{exp2bis}), we find:
$$
\sqrt{q q_0} \;(B^{11}_2({\bf k},{\bf k}_0;{\bf k}-{\bf k}_1)+B^{11}_2({\bf k},{\bf k}_0;{\bf k}_0 +{\bf k}_1))
$$
$$=\frac{1}{2}[{\tilde G}^{(2)}( {\bf k}_1,{\bf k}- {\bf k}_0-{\bf k}_1;{\bf k},{\bf k}_0)+{\tilde G}^{(2)}( {\bf k}- {\bf k}_0-{\bf k}_1, {\bf k}_1;{\bf k},{\bf k}_0)]
$$
$$
=  {\tilde G}^{(2)}( {\bf k}_1,{\bf k}- {\bf k}_0-{\bf k}_1;{\bf k},{\bf k}_0)= A^{(2)}({\bf k},{\bf k}_0) 
$$
\be
-D^{(2)}_{\mu \nu}({\bf k},{\bf k}_0)  \;  {\bf k}_1^{\mu}  ( {\bf k}- {\bf k}_0-{\bf k}_1)^{\nu}\;.
\ee
Therefore
$$
\sqrt{q q_0} \; M^{11}({\bf k},{\bf k}_0;{\bf k}_1)=A^{(2)}({\bf k},{\bf k}_0)-{\rm i}(q+q_0)A^{(1)}({\bf k},{\bf k}_0) 
$$
\be
-D^{(2)}_{\mu \nu}({\bf k},{\bf k}_0)    {\bf k}_1^{\mu}  ( {\bf k}- {\bf k}_0-{\bf k}_1)^{\nu}
= -D^{(2)}_{\mu \nu} ({\bf k},{\bf k}_0)    {\bf k}_1^{\mu}  ( {\bf k}- {\bf k}_0-{\bf k}_1)^{\nu}\;,
\ee
where in the last passage we used the recursive relation Eq. (\ref{rec2}) to cancel the first two terms in the intermediate expression. Then:
$$ 
\sqrt{q q_0}\int d^2 {\bf k}_1 \;M^{11}({\bf k},{\bf k}_0;{\bf k}_1) {\tilde h}({\bf k}_1)\; e^{{\rm i} {\bf k}_1 \cdot {\bf x}}
$$
$$
=-D^{(2)}_{\mu \nu}({\bf k},{\bf k}_0)  \int d^2 {\bf k}_1  \;e^{{\rm i} {\bf k}_1 \cdot {\bf x}}   \,{\bf k}_1^{\mu}  ( {\bf k}- {\bf k}_0-{\bf k}_1)^{\nu}\; {\tilde h}({\bf k}_1)\; e^{{\rm i} {\bf k}_1 \cdot {\bf x}}
$$
\be
= {\rm i }\, D^{(2)}_{\mu \nu}({\bf k},{\bf k}_0)  ( {\bf k}- {\bf k}_0+{\rm i} \,\partial)_{\nu} \partial_{\mu} h({\bf x})\;.
\ee
Using the above formula, we see that the term of Eq. (\ref{ssavor}) involving the ${\bf k}_1$ integral can be recast as: 
$$
-\frac{{\rm i}\sqrt{q q_0}}{2(q+q_0)}\int d^2{\bf x} \;e^{{\rm i}[({\bf k}_0-{\bf k}) \cdot {\bf x}+(q+q_0)h({\bf x})]}  
$$
$$
\times\int d^2 {\bf k}_1 \;M^{11}({\bf k},{\bf k}_0;{\bf k}_1) {\tilde h}({\bf k}_1)\; e^{{\rm i} {\bf k}_1 \cdot {\bf x}}
$$
$$
=\frac{D^{(2)}_{\mu \nu}({\bf k},{\bf k}_0)  }{2(q+q_0)} \int d^2{\bf x} \;e^{{\rm i}[({\bf k}_0-{\bf k}) \cdot {\bf x}+(q+q_0)h({\bf x})]} 
$$
$$ \times \, ( {\bf k}- {\bf k}_0+{\rm i} \,\partial)^{\nu} \partial_{\mu} h({\bf x})
$$
\be
=\frac{1}{2}D^{(2)}_{\mu \nu}({\bf k},{\bf k}_0) \int d^2{\bf x} \;e^{{\rm i}[({\bf k}_0-{\bf k}) \cdot {\bf x}+(q+q_0)h({\bf x})]} \partial_{\mu} h({\bf x}) \partial_{\nu} h({\bf x})\;, 
\ee
where in the last passage we performed an integration by parts on the term involving second derivatives of $h$. Comparison with Eq. (\ref{ssa2}) shows that the above term coincides with the second term on the r.h.s of Eq. (\ref{ssa2}).

 \section{Conclusions}

We have shown that the SA describing scattering of a wave by a gently curved surface  admits a  derivative expansion in powers of derivatives of the height profile. This derivative expansion of the SA has been derived here by performing an  infinite resummation of the perturbative series for the SA, to all orders in the amplitude of the corrugation. Based on this formal derivation it can be expected that the derivative expansion is asymptotically exact in the limit of quasi-specular scattering.  In the leading order the derivative expansion coincides with the classic KA, while in the next  order it provides the leading  curvature correction to the KA. We have also shown that the derivative expansion  is equivalent to the order   $(\nabla h)^2$ to
 the SSA {\it ansatz}, proposed some time ago by Voronovich to describe wave scattering by a  rough surface.  The resummation of the perturbative series performed here to order $(\nabla h)^2$, can be easily generalized to higher orders in $\nabla h$, provided only that the perturbative kernels for the scattering amplitude admit a Taylor expansion of sufficiently high order for small in-plane wave vectors.   

\section{Acnowledgment}

The author thanks T. Emig, N. Graham, M. Kruger,  R. L. Jaffe and M. Kardar for valuable discussions while the manuscript was in preparation.

\end{document}